\documentstyle[12pt]{article}

\makeatletter
\def\@cite#1#2{$^{\hbox{\scriptsize{#1\if@tempswa , #2\fi}}}$}
\makeatother
\def\thefootnote{\fnsymbol{footnote}}

\newlength{\minitwocolumn}
\catcode`\@=11
\long\def\@makefntext#1{
\protect\noindent \hbox to 3.2pt {\hskip-.9pt  
$^{{\eightrm\@thefnmark}}$\hfil}#1\hfill}               

\def\thefootnote{\fnsymbol{footnote}}
\def\@makefnmark{\hbox to 0pt{$^{\@thefnmark}$\hss}}    
        
\def\ps@myheadings{\let\@mkboth\@gobbletwo
\def\@oddhead{\hbox{}
\rightmark\hfil\eightrm\thepage}   
\def\@oddfoot{}\def\@evenhead{\eightrm\thepage\hfil
\leftmark\hbox{}}\def\@evenfoot{}
\def\sectionmark##1{}\def\subsectionmark##1{}}

\font\eightrm=cmr8

\font\sc=cmr5 scaled\magstep1

\def\PL{Phys.~Lett. }

\def\brs{{\delta_{\hbox{\sc B}}}}

\def\delzero{\delta_0}
\def\brs{\delta}

\def\ch{{c}}
\def\ca{{t}}

\def\vv{{v}}

\def\delh{{\delta_1}}
\def\dela{{\delta_2}}
\def\delb{{\delta_3}}
\def\eph{{\epsilon_1}}
\def\epa{{\epsilon_2}}
\def\epb{{\epsilon_3}}

\newcommand\gh{{\rm gh}}
\newcommand{\lb}[2]{[\![#1\,,#2]\!]}
\newcommand{\rd}{\overleftarrow{\partial}} 
\newcommand{\ld}{\overrightarrow{\partial}} 
\newcommand{\sbv}[2]{{\left(\!\left({\,{#1}\,,\,{#2}\,}\right)\!\right)}}
\newcommand{\bracket}[2]{\langle #1\,,#2\rangle}

\def\ba{\mbox{\boldmath $A$}}
\def\bb{\mbox{\boldmath $B$}}

\def\bphi{\mbox{\boldmath $\phi$}}

\newcommand{\calE}{{\cal E}}
\newcommand{\calM}{{\cal M}}

\begin{document}


\baselineskip 0.7cm

\begin{titlepage}
\begin{flushright}
\end{flushright}

\vskip 1.35cm
\begin{center}
{\Large \bf
Chern-Simons Gauge Theory coupled with BF Theory
}
\vskip 1.2cm
Noriaki IKEDA
\footnote{ E-mail address:\ ikeda@yukawa.kyoto-u.ac.jp}
\vskip 0.4cm
{\it Ritsumeikan University \\
Kusatsu, Shiga 525-8577, Japan }\\
and \\
{\it Setsunan University \\
Neyagawa, Osaka 572-8508, Japan }
\date{}

\vskip 1.5cm

\begin{abstract}
We couple three-dimensional Chern-Simons gauge theory with BF theory
and study deformations of the theory by
means of the antifield BRST formalism.
We analyze all possible consistent interaction terms for the action
under physical requirements and find a new topological field theory in
three dimensions with new nontrivial terms and a nontrivial gauge
symmetry. 
We analyze the gauge symmetry of the theory and point out the
theory has the gauge symmetry based on the Courant algebroid.
\end{abstract}
\end{center}
\end{titlepage}

\renewcommand{\thefootnote}{\alph{footnote}}

\setcounter{page}{2}


\rm
\section{Introduction}
\noindent
The Chern-Simons gauge theory in three dimensions is a Schwarz type
topological field theory \cite{BBRT}.
In this paper, we analyze nontrivial deformations of the
Chern-Simons gauge theory in three dimensions as a topological field
theory by the deformation theory of gauge symmetry.

The Chern-Simons gauge theory appears in many scenes of mathematics and
physics.
One of the main applications of the Chern-Simons gauge theory is to
the knot theory \cite{Witten:1989hf}.
The connections of the Chern-Simons gauge theory to 
several knot and link invariants are reviewed in
[3].
The Einstein-Hilbert action in three dimensional gravitational theory
can be formulated as a Chern-Simons gauge theory \cite{Wit}.
In the cubic string field theory, 
the action has the integral of the Chern-Simons type three-form
\cite{Witten:1986cc, WCS}.
[7] reviews the Chern-Simons gauge theory and its
applications.
The purterbation theory has been discussed in [8]. 

Gauge symmetry is one of the fundamental principles of the quantum
field theory.
A deformation theory of the gauge theory \cite{BH, BBH} is a
powerful method to construct a new gauge theory or to prove
impossibility of the construction of new gauge theories under a
certain condition. 
We can construct gauge theories with generalized gauge algebras
by this method.
'Generalized' means that the gauge algebra of the theory 
are not based on usual Lie groups but based on an extended algebra as
a constraint system.
In general, 'structure constants' depend on fields and
are structure functions.
Moreover in this case the gauge algebra is usually the open algebra,
therefore we need analyze the gauge theory by the
Batalin-Vilkovisky (antifield BRST) formalism.

The Chern-Simons gauge theory is constructed from a $1$-form gauge
field $A^a$.
The action of the abelian Chern-Simons theory is as follows:
\begin{eqnarray}
S_{{\sc ACS}} =  \int_{X} \frac{k_{ab}}{2} A^a \wedge d A^b,
\label{acs}
\end{eqnarray}
where $k_{ab}$ is a symmetric constant tensor and
$X$ is a three dimensional manifold. 
Of course, this theory has the abelian gauge symmetry,
$\delzero A^a = d c^a,$
where $c^a$ is a gauge parameter.

Barnich and Henneaux have proved that we can only deform this theory
to the known non-abelian Chern-Simons gauge theory from the
consistency of the gauge symmetry and locality of the action \cite{BH}.
That is, the only consistent gauge theory which we can obtain 
as deformations of the action (\ref{acs}) is the nonabelian Chern-Simons
gauge theory: 
\begin{eqnarray}
S_{{\sc CS}} =  \int_{X} \left( \frac{k_{ab}}{2} A^a \wedge d A^b
+ \frac{1}{6} f_{abc} A^a \wedge A^b \wedge A^c \right), 
\label{nacs}
\end{eqnarray}
where $f_{abc}$ satisfies the relation of the structure constants of the
Lie algebra. 

In the string field theory, we can generalize the cubic string
field theory to the nonpolynomial string field theory with extended
gauge algebras,
$A_{\infty}$-- or $L_{\infty}$--algebra structures 
\cite{Zwiebach:1998fe}\cite{Nakatsu:2001da}\cite{Kajiura:2001ng}.
It seems to be natural if we can deform the Chern-Simons gauge theory 
to a field theory with extended gauge algebras.
However it is only analogical motivation, and 
relation of our theory with string field theory is 
out of scope in this paper.

A generalization of the Chern-Simons gauge theory has also been
discussed in [15] or [16].
We consider an other generalization in this paper.
Now we can couple the Chern-Simons gauge theory 
with an another Schwarz-type topological field theory, BF theory.
We call this theory as the Chern-Simons-BF theory.
Then we systematically analyze all the BRST cohomologies and possible
deformations.
We find a nontrivial new deformation of the gauge symmetry and a
new action.
We can consider that this theory is a higher dimensional
generalization of the nonlinear gauge theory\cite{II1, SS}.


This paper is organized as follows.
In section 2, we construct the superfield antifield formalism
of the abelian Chern-Simons BF theory.
In section 3, we analyze deformations of the abelian Chern-Simons BF
theory and obtain all possible deformations.
In section 4, we calculate the explicit action of our theory.
In section 5, we analyze the gauge symmetry of the theory and find
that the gauge symmetry has the Courant algebroid structure.
Section 6 is conclusion and discussion.

\section{Superfield Formalism of the Abelian Chern-Simons-BF Theory}
\noindent
We begin with a three dimensional base manifold $X$, a target manifold 
$M$ in $N$ dimensions and smooth maps $\phi:X\to M$  with local
coordinate expression $\{\phi^i\}$.
We also have a vector bundle $E$ over $X$.

In three dimensions, abelian Chern-Simons Theory 
with the abelian BF theory has the following action:
\begin{eqnarray}
S_{{\hbox{\sc A}}} 
=  \int_{X} \left( \frac{k_{ab}}{2} A^a \wedge d A^b - B_i \wedge d
\phi^i
\right), 
\label{abf}
\end{eqnarray}
where 
$\phi^i$ is a 0-form scalar field, 
$A^a$ is a $1$-form and $B_i$ is a $2$-form gauge field and
$k_{ab}$ is a symmetric constant tensor.
We call 
the spacetime integration of a BF term as BF theories, where
A is a $p$-form, F is a curvature of A and B is a $n-p-1$-form
in $n$ dimensions.
In three dimensions, there are two actions, the $p=1$ action and the
$p=0$ action. The $p=2$ action is equivalent to the $p=0$ action if
the term is integrated by parts.

The sign factor $-1$ before the second term is introduced for convenience.
We assume that $k_{ab}$ is nondegenerate and has an inverse.
However it is not necessary that $k_{ab}$ is positive definite.

We can take different target spaces for the first term and
the second term in the action (\ref{abf})\cite{TS}.
Indices $a, b, c$, represent indices on the fiber of $E$ and 
and $i, j, k$, represent indices on M, the tangent and cotangent space
of $M$.
%

%
We can add the following usual BF term to the action as a topological
field theory:
\begin{eqnarray}
\int C_a \wedge d A^a, 
\label{abf2}
\end{eqnarray}
where $C_a$
is an auxiliary $1$-form field. 
The action still have the abelian gauge symmetry. 
However if we make the local field redefinition 
$C_a{}^\prime = C_a + \frac{1}{2} k_{ab} A^b$,
the theory reduces to the pure abelian BF theory, 
which deformation is already discussed in the papers 
[20].

We can consider the more general terms 
$(k_{ab}(\phi)/2) \ A^a \wedge d A^b$ or $m^i{}_j(\phi) B_i \wedge 
d \phi^j$ in the action, where $k_{ab}(\phi)$ and $m^i{}_j(\phi)$ are
functions of $\phi^i$. 
However these terms reduces to
the action (\ref{abf}) by local field redefinitions.
If two actions coincide by a local redefinition 
of fields, two theories are
equivalent at least classically. 
We call the theory with the action (\ref{abf}) 
the abelian Chern-Simons-BF theory.

This action has the following abelian gauge symmetry:
\begin{eqnarray}
&& \delzero \phi^i = 0,  \qquad \delzero A^a = d c^a, 
\qquad \delzero c^a = 0,
\nonumber \\
&& \delzero B_i = d t_i, \qquad \delzero t_i = d \vv_i, 
\qquad \delzero \vv_i = 0,
\label{abrs}
\end{eqnarray}
where $c^a$ is a 0-form gauge parameter 
and $t_i$ is a 1-form gauge parameter.
Since $B_i$ is 2-form, we need a 'ghost for ghost' 0-form $\vv_i$.

In order to analyze the theory by the antifield BRST formalism, 
first we take $\ch^a$ and $\ca_i$ to be the Grassmann odd
FP ghosts with ghost number one, and $\vv_i$ to be a the Grassmann even
ghost with ghost number two.
Next we introduce the antifields for all the fields.
Let $\Phi^+$ denote the antifields for the field $\Phi$.
Note that the relations ${\rm deg}(\Phi) + {\rm deg}(\Phi^+) = 3$ and
${\rm gh}(\Phi) + {\rm gh}(\Phi^+) = -1$ are required,
where ${\rm deg}(\Phi)$ and ${\rm deg}(\Phi^+)$ are the form degrees 
of the fields $\Phi$ and $\Phi^+$
and ${\rm gh}(\Phi)$ and ${\rm gh}(\Phi^+)$ are the ghost numbers of
them.
For functions $F(\Phi, \Phi^+)$ and $G(\Phi, \Phi^+)$ of the fields
and the antifields,
we define the antibracket as follows;
\begin{eqnarray}
(F, G) \equiv \frac{F \rd}{\partial \Phi} \frac{\ld G}{\partial \Phi^+}
- 
\frac{F \rd}{\partial \Phi^+} \frac{\ld G}{\partial \Phi},
\label{bracket}
\end{eqnarray}
where ${\rd}/{\partial \varphi}$ and ${\ld}/{\partial
\varphi}$ are the right differentiation and the left differentiation
with respect to $\varphi$, respectively.
If $S, T$ are two functionals, the antibracket is defined as follows:
\begin{eqnarray}
(S, T) \equiv \int_{X}
\left(
\frac{S \rd}{\partial \Phi} \frac{\ld T}{\partial \Phi^+}
- 
\frac{S \rd}{\partial \Phi^+} \frac{\ld T}{\partial \Phi}.
\right)
\label{antibracket}
\end{eqnarray}

The Batalin-Vilkovisky action with the antifields is constructed as
follows:
\begin{eqnarray}
S_0 =  \int_{X} \left( \frac{k_{ab}}{2} A^a \wedge d A^b 
- B_i \wedge d \phi^i
- A^{+}_a \wedge d c^a + B^{+i} \wedge d t_i + t^{+i} \wedge d v_i
\right).
\label{bvaction}
\end{eqnarray}
The gauge transformation is defined as $\delzero F = (S_0, F)$ in the
BV action.
Then the action (\ref{bvaction}) has the gauge transformation 
(the BRST transformation) (\ref{abrs}).
The BRST transformation on all fields are calculated as follows:
\begin{eqnarray}
&& \delzero c^+_a= d A^+_a, \qquad
\delzero A^+_a = k_{ab} d A^b, \nonumber \\
&& \delzero A^a = - d c^a, \qquad
\delzero c^a = 0, \nonumber \\
&& \delzero v^{+i} = - d t^{+i}, \qquad
\delzero t^{+i} = d B^{+i}, \nonumber \\
&& \delzero B^{+i} = - d \phi^i, \qquad
\delzero \phi^i = 0, \nonumber \\
&& \delzero \phi^+_i = d B_i, \qquad
\delzero B_i = d t_i, \nonumber \\
&& \delzero t_i = - d v_i, \qquad
\delzero v_i   = 0, 
\label{abelgauge}
\end{eqnarray}

In order to simplify 
notations and calculations, we rewrite notations by the
superfield formalism.
%
We combine the field, its antifield and their gauge descendant fields
as superfield components. 
%
For $\phi^i$, $A^{a}$ and $B_i$, 
we define corresponding superfields as follows:
\begin{eqnarray}
\bphi^i &= &
\phi^i + B^{+i} + t^{+i} + v^{+i},
\nonumber \\
\ba^a &=& 
c^{a} + A{}^{a} + k^{ab} A^+_b + k^{ab} c^+_b, 
\nonumber \\
\bb_{i} &= &
v_i + t_i + B_i + \phi^+_i.
\label{component}
\end{eqnarray}
Then we define the total degree $|F| \equiv \gh F + \deg F$.
The component fields in a superfield have the same total degree.
The total degrees of $\bphi^i$, 
$\ba^a$ and $\bb_i$ are $0$, $1$ and $2$, respectively.

We introduce a notation $\cdot$ as
the {\it dot product} among superfields in order to simplify the sign 
factors \cite{Cattaneo:2000mc}.
The definitions and properties of the {\it dot product} are listed in
the appendix B. 

The antibracket (\ref{bracket}) and (\ref{antibracket}) are rewritten 
to the {\it dot antibracket} on superfields and {\it dot product}.
The {\it dot antibracket} of the superfields $F$ and $G$ is defined as
\begin{eqnarray}
\sbv{F}{G} \equiv (-1)^{(\gh F + 1) (\deg G - 3)} 
(-1)^{\gh \Phi (\deg \Phi - 3) + 3} (F, G), 
\label{dotantibradef}
\end{eqnarray}
The properties are listed in the appendix B.
we can rewrite the BV antibracket on two superfields $F$ and $G$ 
from (\ref{bracket}) and (\ref{dotantibra}) as follows:
\begin{eqnarray}
\sbv{F}{G} \equiv 
F \cdot \frac{\rd}{\partial \ba^a} \cdot k^{ab} 
\frac{\ld }{\partial \ba^b} \cdot G
+ 
F \cdot \frac{\rd}{\partial \bphi^i} \cdot
\frac{\ld }{\partial \bb_i} \cdot G
- 
F \cdot \frac{\rd }{\partial \bb_{i}} \cdot
\frac{\ld }{\partial \bphi^i} \cdot G.
\label{bvbracket}
\end{eqnarray}
%

We rewrite the Batalin-Vilkovisky action (\ref{bvaction})
for the abelian Chern-Simons-BF theory
by the superfields as follows:
\begin{eqnarray}
S_0 = \int_{X} \left( \frac{k_{ab}}{2} \ba^a \cdot d \ba^b
- \bb_i \cdot d \bphi^i
\right), 
\label{sabf}
\end{eqnarray}
where we integrate only $3$-form part of the integrand.
Integration on $X$ is always understood as the integration of the
$3$-form part of the integrand.
%
The BRST transformation for a superfield $F$ under the
action above is obtained as
\begin{eqnarray}
\delzero F = \sbv{S_0}{F}=
S_0 \cdot \frac{\rd}{\partial \ba^a} \cdot k^{ab} 
\frac{\ld }{\partial \ba^b} \cdot F
+ 
S_0 \cdot \frac{\rd}{\partial \bphi^i} \cdot
\frac{\ld }{\partial \bb_i} \cdot F
- 
S_0 \cdot \frac{\rd }{\partial \bb_{i}} \cdot
\frac{\ld }{\partial \bphi^i} \cdot F.
\end{eqnarray}
Hence we can summarize the BRST transformations on $\bphi^i$,
$\ba^a$ and $\bb_{i}$ as follows:
\begin{eqnarray}
\delzero \bphi^a
&=& \sbv{S_0}{\bphi^i}= d \bphi^i, 
\nonumber \\
\delzero \ba^a
&=& \sbv{S_0}{\ba^a}= d \ba^a, 
\nonumber \\
\delzero \bb_a
&=& \sbv{S_0}{\bb_{i}}= d \bb_{i},
\label{ababelBRST}
\end{eqnarray}
which coincide with (\ref{abelgauge}) 
if we expand them to the component fields.
Equations of motion are
\begin{eqnarray}
d \bphi^i = 0, 
\qquad 
d \ba^a = 0, 
\qquad 
d \bb_{i} = 0.
\label{aeqom}
\end{eqnarray}


$S_0$ must be BRST invariant. In fact, 
\begin{eqnarray}
\delzero S_0 = \sbv{S_0}{S_0}
&=& 2 \int_{X}
d \left(\frac{k^{ab}}{2} \ba^a \cdot d \ba^b
- \bb_{i} \cdot d \bphi^i \right) \nonumber \\
&=& 2 \int_{X}
d \left(\frac{k^{ab}}{2} \ba^a \cdot d \ba^b
- d \bb_{i} \cdot \bphi^i \right),
\label{zerobrs}
\end{eqnarray}
therefore if the base manifold $X$ has no boundary, 
simply $\delzero S_0 = 0$.
If $X$ has a boundary 
we can take two kinds of boundary conditions 
(i) ${\ba^a}_{//}|_{\partial X} = 0$ and 
${\bb_{i}}_{//}|_{\partial X} = 0$, or 
(i) ${\ba^a}_{//}|_{\partial X} = 0$ and 
${\bphi^i}|_{\partial X} = 0$,
where the notation ${//}$ mean the components 
along the direction tangent to the boundary $\partial X$.
We can also take different boundary conditions on each field
component so as to satisfy BRST invariant condition
of the action.
In the rest of this paper, we select appropriate boundary conditions
so as to satisfy $\delzero S_0 = 0$ if we consider $X$ with
boundaries. 
%

\section{Deformation of Chern-Simons-BF Theory}
\noindent
Let us consider a deformation of the action $S_0$
perturbatively,
\begin{eqnarray}
&& S = S_0 + g S_1 + g^2 S_2 + \cdots,
\label{pertur}
\end{eqnarray}
where $g$ is a deformation parameter, or a coupling constant of the
theory. 

In order for the deformed BRST transformation $\brs$ to be nilpotent
and make the theory consistent,
the total action $S$ has to satisfy the following classical master
equation:
\begin{eqnarray}
\sbv{S}{S} = 0.
\label{master}
\end{eqnarray}
Substituting (\ref{pertur}) to (\ref{master}), we obtain the $g$ power 
expansion of the master equation:
\begin{eqnarray}
\sbv{S}{S} = 
\sbv{S_0}{S_0} 
+ 2g \sbv{S_0}{S_1}
+ g^2 [ \sbv{S_1}{S_1} + 2 \sbv{S_0}{S_2} ] + O(g^3) = 0.
\label{purmaster}
\end{eqnarray}
We solve this equation order by order.
Here we make the physical requirements for the solutions.
We require the Lorentz invariance (Lorentzian case),
or $SO(3)$ invariance (Euclidean case)
of the action.
We assume that $S$ is {\it local}.
This means that $S$ is given by the integration of a {\it local}
Lagrangian,
$
S = \int_{X} {\cal L}.
$
Furthermore we exclude the solution which is the BRST transformation
is not deformed, for example, $\brs=\delzero$, as a trivial one.
This condition is realized by the assumption that each term contains
at least one antifield for $S_i$, where $i \geq 1$.

At the $0$-th order, we obtain $\delzero S_0 = \sbv{S_0}{S_0} =0$, which
is already satisfied from (\ref{zerobrs}).
At the first order of $g$ in the Eq.~(\ref{purmaster}), 
\begin{eqnarray}
\delzero S_1 = \sbv{S_0}{S_1} =0,
\label{1brst}
\end{eqnarray}
is required. 
$S_1$ is given by the integration of a {\it local}
Lagrangian from the assumption:
\begin{eqnarray}
&&
S_1 = \int_{M} {\cal L}_1,
\label{s1act} 
\end{eqnarray}
where ${\cal L}_1$ can be constructed from the superfields 
$\bphi^i$, $\ba^a$ and $\bb_{i}$.
If a monomial in ${\cal L}_1$ includes a differentiation $d$, 
its term is proportional to the equations of motion (\ref{aeqom}).
Therefore its term can be absorbed to the abelian action
(\ref{sabf}) by the local field redefinitions of $\bphi^i$, $\ba^a$
or $\bb_{i}$, and these terms are BRST trivial at the BRST
cohomology \cite{BH}.
%
Hence the nontrivial deformation terms must not include
the differentiation $d$ and 
we can write the candidate ${\cal L}_1$ as 
\begin{eqnarray}
&& S_1 = \int_{X} {\cal L}_1, \nonumber \\
&& {\cal L}_1 
= \sum_{k, l}
F_{kl,\ a_1 \cdots a_k}
{}^{i_1 \cdots i_l}(\bphi)
\cdot \ba^{a_1} \cdots \ba^{a_k}
\cdot \bb_{i_1} \cdots \bb_{i_l},
\label{s1}
\end{eqnarray}
where $F_{kl,\ a_1 \cdots a_k} 
{}^{i_1 \cdots i_l}(\bphi)$ is a function of $\bphi^i$.
In order to consider the general deformations,
we do not require the total degree of ${\cal L}_1$ is $3$.
If the total degree of ${\cal L}_1$ is not $3$, the action $S_1$
includes a nonzero ghost number term. 
Then (\ref{1brst}) is calculated as follows:
\begin{eqnarray}
\delzero S_1 &=& 
\sum_{k, l} \int_{X}
[d F_{kl,\ a_1 \cdots a_k} 
{}^{i_1 \cdots i_l}(\bphi)
\cdot \ba^{a_1} \cdots \ba^{a_k}
\cdot \bb_{i_1} \cdots \bb_{i_l} \nonumber \\
&& + 
\sum_{r=1}^k (-1)^{r-1}
F_{kl,\ a_1 \cdots a_k} 
{}^{i_1 \cdots i_l}(\bphi)
\cdot \ba^{a_1} \cdots d \ba{}^{a_r} \cdots \ba^{a_k}
\cdot \bb_{i_1} \cdots \bb_{i_l} \nonumber \\
&& + 
\sum_{s=1}^l (-1)^{k}
F_{kl,\ a_1 \cdots a_k} 
{}^{i_1 \cdots i_l}(\bphi)
\cdot \ba^{a_1} \cdots \ba^{a_k}
\cdot \bb_{i_1} \cdots d \bb_{i_s} \cdots \bb_{i_l}]
\nonumber \\
%
&=& 
\sum_{k, l} \int_{X}
d [F_{kl, \ a_1 \cdots a_k} 
{}^{i_1 \cdots i_l}(\bphi^a)
\cdot \ba^{a_1} \cdots \ba^{a_k}
\cdot \bb_{i_1} \cdots \bb_{i_l}].
\end{eqnarray}
If there is no boundary in $X$, there is no restriction for $S_1$ and 
we obtain $\delzero S_1 = 0$.
If there are boundaries in $X$, $\delzero S_1 = 0$
if 
\begin{eqnarray}
(F_{kl, \ a_1 \cdots a_k} 
{}^{i_1 \cdots i_l}(\bphi^a)
\cdot \ba^{a_1} \cdots \ba^{a_k}
\cdot \bb_{i_1} \cdots \bb_{i_l})_{//}|_{\partial X} = 0.
\label{s1boundary}
\end{eqnarray}
$S_1$ must be constructed from the terms which satisfy the
requirements above.
If we take the boundary condition (i)
then (\ref{s1boundary}) is satisfied if 
the terms include at least one ${\ba^a}$ or one ${\bb_i}$ . 
If we take (ii) 
then (\ref{s1boundary}) is satisfied if 
the terms include at least one ${\ba^a}$ or one ${\bphi^i}$. 

At the second order of $g$, 
\begin{eqnarray}
\sbv{S_1}{S_1} + 2 \sbv{S_0}{S_2} = 0, 
\label{pursec}
\end{eqnarray}
is required.
We cannot construct nontrivial $S_2$ to satisfy (\ref{pursec})
from the integration of a local Lagrangian, because
$\delzero$-BRST transforms of the local terms are always total
derivative.
Therefore if we assume locality of the action, $S_2$ is BRST
trivial (the Poincar\'e lemma), because we consider the local
deformations on the space of field theories.
If we solve the higher order $g$ part of the 
equation (\ref{purmaster}) recursively,
we can find that $S_\alpha$ is BRST trivial for $\alpha \geq 2$.
Therefore we can set $S_\alpha = 0$ for $\alpha \geq 2$. 
%
Then the condition (\ref{pursec}) reduces to
\begin{eqnarray}
\sbv{S_1}{S_1} = 0.
\label{s1s1}
\end{eqnarray}
This equation imposes the identities on the structure functions 
$F_{kl, \ a_1 \cdots a_k} {}^{i_1 \cdots
i_l}(\bphi)$ in (\ref{s1}).
Now we have obtained the possible deformations of 
the Chern-Simons-BF theory in three
dimensions from (\ref{sabf}) and (\ref{s1}) as follows:
\begin{eqnarray}
S = S_0 + g S_1 
&=& \int_{X} \biggl( \frac{k_{ab}}{2} \ba^a \cdot d \ba^b 
- \bb_i \cdot d \bphi^i \nonumber \\
&& + g \sum_{k, l}
F_{kl,\ a_1 \cdots a_k}
{}^{i_1 \cdots i_l}(\bphi)
\cdot \ba^{a_1} \cdots \ba^{a_k}
\cdot \bb_{i_1} \cdots \bb_{i_l} \biggl),
\label{lapS}
\end{eqnarray}
with the condition (\ref{s1s1}) on the structure functions 
$F_{kl,\ a_1 \cdots a_k}{}^{i_1 \cdots i_l}(\bphi)$.
The master equation (\ref{master}) reduces to 
$\delzero S_1 + g/2 \sbv{S_1}{S_1} = 0$.
This is nothing but the Maurer-Cartan equation under the differential
$\delzero$.


\def\delh{{\delta_1}}
\def\dela{{\delta_3}}
\def\delb{{\delta_2}}
\def\eph{{\epsilon_1}}
\def\epa{{\epsilon_3}}
\def\epb{{\epsilon_2}}
\def\tx{{\cal X}}
\def\qstr{{\hat Q}}

\section{Chern-Simons Sigma Model}
As a nontrivial example, let us solve the condition (\ref{s1s1})
explicitly in case that the ghost number of the total action is zero.
This assumption enables us to restrict the action to the following
form:
\begin{eqnarray}
S = \int_{X} \left( \frac{k_{ab}}{2} \ba^a \cdot d \ba^b 
- \bb_i \cdot d \bphi^i 
+  f_{1a}{}^i (\bphi) \cdot \ba^a \cdot \bb_i 
+  \frac{1}{6} f_{2abc} (\bphi) \cdot \ba^a \cdot \ba^b \cdot \ba^c
\right),
\label{gzaction}
\end{eqnarray}
where we rewrite two structure functions $f_{1a}{}^i = g F_{11, a}{}^i$
and $\frac{1}{6} f_{2abc} = g F_{30, abc}$ for clarity.

If we substitute (\ref{gzaction}) to the condition (\ref{s1s1}),
we obtain the identities on the structure functions 
$f_{1a}{}^i$ and $f_{2abc}$ as follows:
\begin{eqnarray}
&& k^{ab} f_{1a}{}^i \cdot f_{1b}{}^j = 0, \nonumber \\ 
&& \frac{\partial f_{1b}{}^i}{\partial \bphi^j} \cdot f_{1c}{}^j
- \frac{\partial f_{1c}{}^i}{\partial \bphi^j} \cdot f_{1b}{}^j
+ k^{ef} f_{1e}{}^i \cdot f_{2fbc} = 0, \nonumber \\
&& \left( f_{1d}{}^j \cdot \frac{\partial f_{2abc}}{\partial \bphi^j}
- f_{1c}{}^j \cdot \frac{\partial f_{2dab}}{\partial \bphi^j}
+ f_{1b}{}^j \cdot \frac{\partial f_{2cda}}{\partial \bphi^j}
- f_{1a}{}^j \cdot \frac{\partial f_{2bcd}}{\partial \bphi^j} 
\right) \nonumber \\
&& 
+ k^{ef} (f_{2eab} \cdot f_{2cdf} 
+ f_{2eac} \cdot f_{2dbf} 
+ f_{2ead} \cdot f_{2bcf})
= 0.
\label{stride}
\end{eqnarray}
The BRST transformation of each field is calculated from 
the definition of the BRST transformation 
$\brs F = \sbv{S}{F}$:
\begin{eqnarray}
&& \brs \ba^a = d \ba^a +  k^{ab} f_{1b}{}^j \cdot \bb_j
+  \frac{1}{2} k^{ab} f_{2bcd} \cdot \ba^c \cdot \ba^d, \nonumber \\
&& \brs \bb_i =d \bb_i
+  \frac{\partial f_{1b}{}^j}{\partial \bphi^i} \cdot \ba^b \cdot \bb_j
+  \frac{1}{6} \frac{\partial f_{2bcd}}{\partial \bphi^i} 
\cdot \ba^b \cdot \ba^c \cdot \ba^d, \nonumber \\
&& \brs \bphi^i = d \bphi^i -  f_{1b}{}^i \cdot \ba^b.
\label{brst}
\end{eqnarray}

If we set all the antifields zero, we obtain the usual action without
antifields as follows:
\begin{eqnarray}
S = \int_{X} \left( \frac{k_{ab}}{2} A^a \wedge d A^b 
- B_i \wedge d \phi^i
+  f_{1a}{}^i (\phi) A^a B_i
+  \frac{1}{6} f_{2abc} (\phi) A^a A^b A^c \right),
\end{eqnarray}
with the gauge symmetry:
\begin{eqnarray}
&& \brs A^a = d c^a +  k^{ab} f_{1b}{}^j t_j
+  k^{ab} f_{2bcd} A^c c^d, \nonumber \\
&& \brs B_i =d t_i
+  \frac{\partial f_{1b}{}^j}{\partial \phi^i} (A^b t_j - c^b B_j)
+  \frac{1}{2} \frac{\partial f_{2bcd}}{\partial \phi^i} A^b A^c c^d, 
\nonumber \\
&& \brs \phi^i = - f_{1b}{}^i c^b.
\end{eqnarray}
The identities on the structure functions is obtained as:
\begin{eqnarray}
&& k^{ab} f_{1a}{}^i (\phi) f_{1b}{}^j (\phi) = 0, 
\nonumber \\ 
&& \frac{\partial f_{1b}{}^i (\phi) }{\partial \phi^j} f_{1c}{}^j (\phi) 
- \frac{\partial f_{1c}{}^i (\phi) }{\partial \phi^j} f_{1b}{}^j (\phi) 
+ k^{ef} f_{1e}{}^i (\phi) f_{2fbc} (\phi) = 0, 
\nonumber \\
&& \left( f_{1d}{}^j (\phi) \frac{\partial f_{2abc} (\phi)}
{\partial \phi^j}
- f_{1c}{}^j (\phi) \frac{\partial f_{2dab} (\phi) }{\partial \phi^j}
+ f_{1b}{}^j (\phi) \frac{\partial f_{2cda} (\phi) }{\partial \phi^j}
- f_{1a}{}^j (\phi) \frac{\partial f_{2bcd} (\phi) }{\partial \phi^j} 
\right) \nonumber \\
&& 
+ k^{ef} (f_{2eab} (\phi) f_{2cdf} (\phi) 
+ f_{2eac} (\phi) f_{2dbf} (\phi) 
+ f_{2ead} (\phi) f_{2bcf} (\phi) )
= 0, 
\label{identity3}
\end{eqnarray}
If $f_{1a}{}^j = 0$ and 
$f_{2abc}$ is a constant, 
(\ref{identity3}) reduces to the usual Jacobi
identity of the Lie algebra structure constants and we have the
nonabelian gauge symmetry.
However in general $f_{2abc} (\phi)$ depends on the fields,
and the theory has a
generalization of the nonabelian gauge symmetry. 


\section{Courant Algebroid}  
\noindent
Let us analyze the identities (\ref{stride}) on the structure
functions $f_1$ and $f_2$, which is equivalent to (\ref{identity3}).
The gauge algebra under this theory is the Courant algebroid.

A Courant algebroid is introduced by Courant 
in order to analyze the Dirac structure as a generalization of the 
Lie algebra of the vector fields
on the vector bundle \cite{Courant, LWX}.
A Courant algebroid is a vector bundle $\calE \rightarrow \calM$
and has a nondegenerate symmetric bilinear form
$\bracket{\cdot}{\cdot}$ 
on the bundle, a bilinear operation $\circ$ on $\Gamma(\calE)$(the space of
sections on $\calE$), an a bundle map 
$\rho: \calE \rightarrow T\calM$ satisfying the following properties
\cite{Roy01}:
\begin{eqnarray}
&& 1, \quad e_1 \circ (e_2 \circ e_3) = (e_1 \circ e_2) \circ e_3 
+ e_2 \circ (e_1 \circ e_3), \nonumber \\
&& 2, \quad \rho(e_1 \circ e_2) = [\rho(e_1), \rho(e_2)], \nonumber \\
&& 3, \quad e_1 \circ F e_2 = F (e_1 \circ e_2)
+ (\rho(e_1)F)e_2, \nonumber \\
&& 4, \quad e_1 \circ e_2 = \frac{1}{2} {\cal D} \bracket{e_1}{e_2},
\nonumber \\ 
&& 5, \quad \rho(e_1) \bracket{e_2}{e_3}
= \bracket{e_1 \circ e_2}{e_3} + \bracket{e_2}{e_1 \circ e_3},
  \label{courantdef}
\end{eqnarray}
where 
$e_1, e_2$ and $e_3$ are sections of $\calE$, $F$ is a function on
$\calM$.
${\cal D}$ is a map from functions on $\calM$ to $\Gamma(\calE)$ and is
defined as 
$\bracket{{\cal D}F}{e} = \rho(e) F$.
Let $e^a$ be basis of $\Gamma(\calE)$ with respect to the fiber.
Then (\ref{courantdef}) is written as
\begin{eqnarray}
&& 1, \quad e^a \circ (e^b \circ e^c) = (e^a \circ e^b) \circ e^c 
+ e^b \circ (e^a \circ e^c), \nonumber \\
&& 2, \quad \rho(e^a \circ e^b) = [\rho(e^a), \rho(e^b)], \nonumber \\
&& 3, \quad e^a \circ F e^b = F (e^a \circ e^b)
+ (\rho(e^a)F)e^b, \nonumber \\
&& 4, \quad e^a \circ e^b = \frac{1}{2} {\cal D} \bracket{e^a}{e^b},
\nonumber \\ 
&& 5, \quad \rho(e^a) \bracket{e^b}{e^c}
= \bracket{e^a \circ e^b}{e^c} + \bracket{e^b}{e^a \circ e^c},
  \label{courantbase}
\end{eqnarray}

Let us consider the supermanifold ${\tilde X}$ which bosonic part is
a three dimensional manifold $X$.
In our topological field theory, 
a base space $\calM$ is the space of a (smooth) map from ${\tilde X}$
to a target space $M$.
Basis on $\Gamma(\calE)$ is $e^a = \ba^a$. 
We define a symmetric bilinear form 
$\bracket{\cdot}{\cdot}$, a bilinear operation $\circ$ and an a
bundle map $\rho$ as follows:
\begin{eqnarray}
&& e^a \circ e^b \equiv \sbv{\sbv{S}{e^a}}{e^b}, \nonumber \\
&& \bracket{e^a}{e^b} \equiv \sbv{e^a}{e^b}, \nonumber \\
&& \rho(e^a) F(\bphi) \equiv \sbv{e^a}{\sbv{S}{F(\bphi)}}, \nonumber \\
&& {\cal D}(*) \equiv \sbv{S}{*}.
  \label{corresbase}
\end{eqnarray}
Then we can easily confirm that the gauge algebra satisfies the
conditions $1$ to $5$ of the Courant algebroid
by the identities (\ref{stride}).

Conversely, first we take the basis on $\ba^a$ on the fiber of
the vector bundle $\calE$.
We define the graded Poisson structure (\ref{bvbracket})
on the bundle $\calE \oplus T^*\calM$,
where the grading on the fiber direction is shifted by $2$.
We define the operations $\bracket{\cdot}{\cdot}$, $\circ$ and
$\rho$ as 
\begin{eqnarray}
&& \ba^a \circ \ba^b = - k^{ac} k^{bd} f_{2cde} (\bphi) \ba^e, 
\nonumber \\
&&\bracket{\ba^a}{\ba^b} = k^{ab},
\nonumber \\
&& \rho(\ba^a) \bphi^i = - f_{1c}{}^i (\bphi) k^{ac}.
  \label{cscourant}
\end{eqnarray}
We can take a Darboux coordinate
such that $\bracket{\ba^a}{\ba^b} = k^{ab}.$.
Then the conditions $1$ to $5$ of the Courant algebroid are equivalent
to the identities (\ref{stride}) on $f_1$ and $f_2$.
%
The action $S$ is the BRST charge of the Courant algebroid.
Since the master equation (\ref{master}) is equivalent to
(\ref{stride}), the relations $1$ to $5$ is represented 
by the master equation of the action $S$.


\section{Conclusion and Discussion}
\noindent
We have considered the Chern-Simons gauge theory
in three dimensions, coupled with the BF theory which is an another  
Schwarz-type topological field theory.
We have analyzed all possible deformations of this theory by the
antifield BRST formalism.  
Then it led us to a deformed new action with a new gauge
symmetry. 
This 'nonlinear' gauge symmetry in our theory
is an extension of the usual Lie algebra and the quantities corresponding
to the structure constants are not constants and functions of the
fields. 

The 'nonlinear' Lie algebras in the nonlinear gauge theory are
recently analyzed in the context of $L_\infty$-algebra 
\cite{LS, Sta, FLS},
or the Lie algebroid \cite{Levin:2000fk, Olshanetsky:2002ur}.
These mathematical notions will be applicable to our theory.
Here we have found that the gauge symmetry of the deformed topological 
field theory constructed in this paper has the gauge symmetry based on 
the Courant algebroid.
Our theory is a first example of field theories with the Courant
algebroid structure.

Since the deformed gauge theory is still a topological field theory,
observables in this theory will define cohomological quantities.
These are regarded as deformations of mathematical invariants
obtained from the Chern-Simons gauge theory.
In the Chern-Simons gauge theory, 
the coupling constant is quantized to the integer variable.
However we have not treated such global aspects in this paper.
The mathematical and physical aspects of this deformation should be
studied. 

We do not analyze the quantum theory in this paper.
Since the gauge algebra in our theory is generally the open algebra,
we have to use the BV formalism
in order to make the gauge fixing and quantize the theory.
We should analyze the correlation functions of the observables
in detail.

\section*{Acknowledgments}
The author would like to thank J.~Stasheff, T.~Strobl and P.~Xu
for valuable comments and discussions.

\section*{Appendix A, Antibracket}
In three dimensions, 
we define the antibracket for functions $F(\Phi, \Phi^+)$ and $G(\Phi,
\Phi^+)$ of the fields and the antifields
as follows;
\begin{eqnarray}
(F, G) \equiv \frac{F \rd}{\partial \Phi} \frac{\ld G}{\partial \Phi^+}
- 
\frac{F \rd}{\partial \Phi^+} \frac{\ld G}{\partial \Phi},
\label{anti}
\end{eqnarray}
where ${\rd}/{\partial \varphi}$ and ${\ld}/{\partial
\varphi}$ are the right differentiation and the left differentiation
with respect to $\varphi$, respectively.
The following identity about left and right derivative is useful:
\begin{eqnarray}
\frac{\ld F}{\partial \varphi} =  (-1)^{(\gh F - \gh \varphi) \gh \varphi 
+ (\deg F - \deg \varphi) \deg \varphi}
\frac{F \rd}{\partial \varphi}.
\label{lrdif}
\end{eqnarray}
If $S, T$ are two functionals, the antibracket is defined as follows:
\begin{eqnarray}
(S, T) \equiv \int_{X}
\left(
\frac{S \rd}{\partial \Phi} \frac{\ld T}{\partial \Phi^+}
- 
\frac{S \rd}{\partial \Phi^+} \frac{\ld T}{\partial \Phi}.
\right)
\label{antif}
\end{eqnarray}
The antibracket satisfies the following identities:
\begin{eqnarray}
&& (F, G) = -(-1)^{(\deg F - 3)(\deg G - 3) + (\gh F + 1)(\gh G + 1)}(G, F),
\nonumber \\
&& (F, GH) = (F, G)H + (-1)^{(\deg F -3)\deg G + (\gh F +1) \gh G} G(F, H),
\nonumber \\
&& (FG, H) = F(G, H) + (-1)^{\deg G(\deg H - 3) + \gh G(\gh H +1) } (F, H)G,
\nonumber \\
&& (-1)^{(\deg F - 3)(\deg H - 3) + (\gh F + 1)(\gh H + 1) } (F, (G, H)) 
+ {\rm cyclic \ permutations} = 0,
\label{antibra}
\end{eqnarray}
where $F, G$ and $H$ are functions on fields and antifields.

\section*{Appendix B, Dot Product}
\noindent
It is convenient to combine fields to superfield to analyze BV
actions.
In order to simplify cumbersome sign factors, we introduce the {\it
dot product}, {\it dot Lie bracket}, {\it dot antibracket} and {\it
dot differential}.

For a superfield $F(\Phi, \Phi^{+})$ and $G(\Phi, \Phi^{+})$,
The following identities are satisfied:
\begin{eqnarray}
&& FG = (-1)^{\gh F \gh G + \deg F \deg G} G F, \nonumber \\
&& d(FG) = dF G + (-1)^{\deg F} F dG, 
\label{FGpro}
\end{eqnarray}
at the usual products.
The graded commutator of two superfields satisfies the following 
identities:
\begin{eqnarray}
&& [F, G] = -(-1)^{\gh F \gh G + \deg F \deg G} [G, F], \nonumber \\
&& [F, [G, H]] = [[F, G], H] 
+ (-1)^{\gh F \gh G + \deg F \deg G} [G, [F, H]].
\label{FGcom}
\end{eqnarray}

We introduce the total degree of a superfield $F$ as 
$|F| = \gh F + \deg F$.
We define the {\it dot product} on superfields as
\begin{eqnarray}
F \cdot G \equiv  (-1)^{\gh F \deg G} FG, 
\label{dotpro}
\end{eqnarray}
and the {\it dot Lie bracket}
\begin{eqnarray}
\lb{F}{G} \equiv (-1)^{\gh F \deg G} [F, G].
\label{dotlie}
\end{eqnarray}
We obtain the following identities
of the {\it dot product} and the {\it dot Lie bracket} from
(\ref{FGpro}), (\ref{FGcom}), (\ref{dotpro}) and (\ref{dotlie}):
\begin{eqnarray}
&& F \cdot G = (-1)^{|F||G|} G \cdot F, \nonumber \\
&& \lb{F}{G} = - (-1)^{|F||G|} \lb{G}{F}, \nonumber \\
&& \lb{F}{\lb{G}{H}} = \lb{\lb{F}{G}}{H}
+ (-1)^{|F||G|} \lb{G}{\lb{F}{H}},
\end{eqnarray}
and
\begin{eqnarray}
d (F \cdot G) \equiv d F \cdot G + (-1)^{|F|} F \cdot d G.
\end{eqnarray}

The {\it dot antibracket} of the superfields $F$ and $G$ is defined as
\begin{eqnarray}
\sbv{F}{G} \equiv (-1)^{(\gh F + 1) (\deg G - 3)} 
(-1)^{\gh \Phi (\deg \Phi - 3) + 3} (F, G), 
\label{dotantibra}
\end{eqnarray}
Then the following identities are obtained from the equations
(\ref{antibra}) and (\ref{dotantibra}):
\begin{eqnarray}
&& \sbv{F}{G} = -(-1)^{|F||G|} \sbv{G}{F},
\nonumber \\
&& \sbv{F}{GH} = \sbv{F}{G} \cdot H 
+ (-1)^{|F||G|} G \cdot \sbv{F}{H},
\nonumber \\
&& \sbv{FG}{H} = F \cdot \sbv{G}{H} 
+ (-1)^{|G||H|} \sbv{F}{H} \cdot G,
\nonumber \\
&& (-1)^{|F||H|} \sbv{F}{\sbv{G}{H}} 
+ {\rm cyclic \ permutations} = 0.
\end{eqnarray}
We define the {\it dot differential} as
\begin{eqnarray}
&& \frac{\ld }{\partial \varphi} \cdot F 
\equiv (-1)^{\gh \varphi \deg F}
\frac{\ld F}{\partial \varphi}, \nonumber \\
&&F \cdot \frac{\rd }{\partial \varphi}
\equiv (-1)^{\gh F \deg \varphi}
\frac{F \rd}{\partial \varphi}.
\end{eqnarray}
Then, ~from the equation (\ref{lrdif}), we can obtain the formula
\begin{eqnarray}
\frac{\ld }{\partial \varphi} \cdot F =   
(-1)^{(|F| - |\varphi|) |\varphi|}
F \cdot \frac{\rd }{\partial \varphi}.
\label{lrdiff}
\end{eqnarray}

\newcommand{\bibit}{\sl}



\vfill\eject
\end{document}